\documentclass[article,prd,10pt,aps,showpacs,showkeys,amssymb,nofootinbib]{revtex4-1}
\usepackage{amsmath,amsfonts,mathrsfs}
\usepackage[dvipsnames]{xcolor}

\newcommand{\nl}{\nonumber\\ }

\newcommand{\vc}[1]{\mathbf{#1}}
\newcommand{\rmd}{{\mathrm d}}
\newcommand{\Vp}{\mathcal{P}}
\newcommand{\sgn}{{\rm sgn}}
\newcommand{\rp}{{\rm Re}}
\newcommand{\ip}{{\rm Im}}
\newcommand{\intvecp}{\int\frac{\mathrm d\mathbf p}{(2\pi)^3}}

\newcommand{\intE}{\int\limits_m^{\Lambda_E}\!\rmd E\,}
\newcommand{\gT}{\mathcal T}
\newcommand{\gX}{\mathcal X}

\newcommand{\gZ}{\mathcal Z}
\newcommand\elimit{\lim_{\epsilon\rightarrow0}}
\newcommand\thfunc[1]{\Theta\!\left(#1\right)}
\newcommand\abs[1]{\left|#1\right|}
\newcommand\vk{\vc{k}}
\newcommand\find[1]{^{(#1)}}
\newcommand\vecprod[2]{#1\times#2}
\newcommand{\etal}{{\it et al.}}

\begin{document}
\title{Calculation of the one loop box integral at Finite Temperature and Density}

\author{A. S. Khvorostukhin}
\email{hvorost@theor.jinr.ru}
\affiliation{Joint Institute for Nuclear Research,
 141980 Dubna, Russia}
\affiliation{ Institute of Applied Physics, MD-2028 Kishineu, Moldova}
\begin{abstract}
Calculation of hadronization, decay or scattering processes at non-zero temperatures and densities within the Nambu-Jona-Lasinio-like models requires some techniques for computation of Feynmann diagrams. Decomposition of Feynman diagrams at the one loop level leads to the appearance of elementary integrals with one, two, three, and four fermion lines.  For example, evaluation of the $\pi\pi$ scattering amplitude requires calculating a box diagram with four fermion lines. In this work, the real and imaginary parts of the box integral at the one loop level are provided in the form suitable for numerical evaluation. The obtained expressions are applicable to any value of temperature, particle mass, and chemical potential. We pay special attention to the conditions for the existence of the appearing improper integrals and correct the results \cite{Klevansky} for the three fermion lines. As a result, we have obtained constraints on possible values of particle momenta. Among the expressions for the box integral, the general formulas for the integral with an arbitrary number of lines are derived for the case with zero or collinear fermion momenta.
\end{abstract}
\pacs{
12.38.Mh,
12.39.-x,
05.90.+m,
05.70.Ce
}
\keywords{NJL model, low energy $\pi\pi$ scattering.}

\maketitle
\section{Introduction}
Due to the complexity of the QCD Lagrangian and the nonperturbativity of QCD at energies relevant for the deconfinement/chiral phase transition, effective theories of strongly interacting matter at non-zero temperatures/densities are under intensive study, see, e.g. \cite{HatsudaKunihiroChiral, FuLiu2009,DBK2013,Moeller2013}, and references cited therein.

Extension of QCD Lagrangian-inspired effective models such as the Nambu-Jona-Lasinio (NJL) one \cite{NJLoriginal, KlevanskyNJL1992, Huefner1995,Zhuang1995,Huefner1994,Rehberg1995} to finite temperatures/densities is performed by the imaginary time Matsubara formalism \cite{Matsubara}. The description of hadronization processes starts with Feynman diagrams which can contain two (a polarization loop), three or four (a box diagram) fermion lines, and the obtained amplitudes at the one loop level are the integrals that contain $\prod_i (q_i^2-m_i^2)^{-1}$. For calculations at zero temperature, it was shown \cite{PassarinoVeltman} that all occurring transition amplitudes at the one loop level can be decomposed into a number of elementary integrals, which in turn can be classified by the number of particle lines they contain. This decomposition technique can be generalized to non-zero temperatures using the Matsubara formalism~\cite{Klevansky}. The authors of the mentioned paper gave a solution of this task for diagrams with one, two, and three fermion lines, explicitly displaying their complex nature.

The study of meson scattering processes (for example, $\pi\pi$ scattering) requires the calculation of a box diagram with four fermion lines \cite{FuLiu2009, JRS2002}.  In the previous works, the amplitude for $\pi\pi$ scattering at non-zero temperatures was calculated only in special kinematics with $s= 4 M_\pi^2$ and $t=u=0$ due to the complexity of calculations in the general case. The aim of this paper is to solve technical problems associated with such integrals following the corrected techniques developed in \cite{Klevansky} and to give researchers, who are active in this field, the knowledge sufficient to calculate such functions. Expressions for both real and imaginary parts will be given in the form suitable for numerical evaluation.

We carefully investigate for what values of the parameters all integrals that appear exist. This work brings us to the necessity of correcting the results \cite{Klevansky} for the three-line elementary integral. We also show that though the integrals with three and more fermion lines seem convergent, our analysis leads to the insistence of using the three momentum cutoff regularization scheme with $|\vc{p}|<\Lambda$.

The paper is organized as follows. Section II shows how the original integral can be represented as a sum of elementary integrals. Section III is devoted to the correction of the result \cite{Klevansky} for the three line function. We underline that one should carefully consider the question of the existence of multiple integrals. After that, the results for the four line integral are given depending on configurations of particle momenta. The simplest cases when all three momenta are collinear or equal to zero are discussed in Sec. IV. More complicated {\it``planar''} configurations are considered in Sec. V. The main original result of the paper is formulated in Sec.~VI, where the expression for the most general case of independent momenta is given. We also shortly discuss how to calculate a one-loop integral with any number of fermion lines in Sec. VII.  Finally, the results are summarized in Conclusions. Some technical details are presented in two Appendices.

\section{Decomposition to elementary integrals}
In the general case, after replacing the integration over the time component of the four-momentum by the sum over the Matsubara frequencies, the integral with $L+1$ fermion lines has the form \cite{Klevansky}:
\begin{align}
\label{Z0def}
&Z_0\find{L}\left(\gX_1^+,\ldots,\gX_L^+;\gT\right)=\frac{16\pi^2}\beta\lim_{\eta\rightarrow0}\sum_ne^{i\omega_n\eta}\intvecp\frac{1}{(i\omega_n+\mu)^2-E^2}
\prod_{l=1}^L
\frac{1}{(i\omega_n+\mu-\lambda_l)^2-\widetilde E_l^2(\vk_l)}\,,
\end{align}
where the arguments were written as a collection of ordered sets (vectors)
\begin{align}
\label{gTdef}
\gX_l^\pm&=(\pm\lambda_l,\vk_l,m_l),\quad \gT=(T,\mu,m).
\end{align}
Here $T$ is the temperature, $\beta=1/T$, $m$ and $m_l$ denote the masses of fermions, $\mu$ and $\mu_l$ are the fermion chemical potentials, $\lambda_l$ is defined as $\lambda_l =\mu-\mu_l+i\nu_{j_l}$, and $\vk_l$ are the 3-momenta. All considered integrals are integrated for $p=|\vc{p}|\leq\Lambda$, i.e. $\Lambda$ is the 3-momentum cutoff, and
\begin{align}
\label{Edef}
E&=\sqrt{p^2+m^2},\quad \widetilde E_l(\vc{k})=\sqrt{(\vc{p}-\vc{k})^2+m_l^2}\,.
\end{align}

The Matsubara summation is carried out over the bosonic $i\nu_{j_l}=2\pi j_lT$ and fermionic $\omega_n=(2n+1)\,\pi\,T$ frequencies. After the Matsubara summation over $n$ is carried out, the complex bosonic frequencies are analytically continued to their values on the real axis and become the zeroth components associated with the corresponding 3-momentum, $i\nu_{j_l}\rightarrow k_l^0$. The limit $\eta\rightarrow0$ has to be taken after the Matsubara summation.

The Matsubara summation in Eq. (\ref{Z0def}) can be easily performed \cite{Klevansky, MatsSum}.  There are  $2(L+1)$ poles in Eq.~(\ref{Z0def})
\begin{align}
i\omega_n&=-\mu\pm E, \quad i\omega_n=\lambda_l-\mu\pm\widetilde E_l(\vc{k}),
\end{align}
and after the summation over $n$, which is performed according to the rule
\begin{equation}
 T \sum_{n}\frac{1}{i\omega_n \pm z} =  \frac{1}{e^{\pm \beta z} + 1} =  f(\pm z)
\end{equation}
with the Fermi-Dirac distribution function $f(\pm z)$, one obtains
\begin{align}
Z_0\find{L}\left(\gX_1^+,\ldots,\gX_L^+;\gT\right)
=16\pi^2&\left[\intvecp\frac{f(E-\mu)}{2E}\prod_{l=1}^L\frac{1}{(E-\lambda_l)^2-\widetilde E_l^2(\vc{k}_l)}\right.
\nl
&-\intvecp\frac{f(-E-\mu)}{2E}\prod_{l=1}^L\frac{1}{(E+\lambda_l)^2-\widetilde E_l^2(\vc{k}_l)}
\nl
&+\sum_{s=1}^L\intvecp\frac{f(E_s-\mu_s)}{2E_s}\frac{1}{(E_s+\lambda_s)^2-\widetilde E^2(\vk_s)}
\prod_{l=1,l\neq s}^L\frac{1}{(E_s+\lambda_{sl})^2-\widetilde E_l^2(\vk_s-\vk_l)}\nl
&\left.-\sum_{s=1}^L\intvecp\frac{f(-E_s-\mu_s)}{2E_s}\frac{1}{(E_s-\lambda_s)^2-\widetilde E^2(\vk_s)}
\prod_{l=1,l\neq s}^L\frac{1}{(E_s-\lambda_{sl})^2-\widetilde E_l^2(\vk_s-\vk_l)}\right].
\end{align}
Above,   the substitution $\vc{p}\rightarrow\vc{k}_s-\vc{p}$  was made in the third and fourth lines, and\footnote{Here and everywhere below $l\neq s$ if we have two indices in the expression.}
\begin{align}
\lambda_{ls}&=\lambda_l-\lambda_s=\mu_s-\mu_l+i\nu_{j_l}-i\nu_{j_s},\quad l\neq s,
\end{align}
\begin{align}
\label{tEldef}
\widetilde E(\vc{k})&=\sqrt{(\vc{p}-\vc{k})^2+m^2},\quad
E_l=\sqrt{p^2+m_l^2}.
\end{align}

So we see that the original integral can be written as a decomposition
\begin{align}
Z_0\find{L}\left(\gX_1^+,\ldots,\gX_L^+;\gT\right)
&=\mathcal{Z}^+_L\left(\gX^-_1,\ldots,\gX_L^-;\gT\right)
-\mathcal{Z}^-_L\left(\gX^+_1,\ldots,\gX_L^+;\gT\right)\nl
&+\sum_{s=1}^L\left[\mathcal{Z}^+_L\left(\gX^+_{s1},\ldots,\gX^+_{sL};\gT_s\right)
-\mathcal{Z}^-_L\left(\gX^-_{s1},\ldots,\gX^-_{sL};\gT_s\right)\right]
\end{align}
with
\begin{align}
\gT_s&=(T,\mu_s,m_s),\quad \gX_{ss}^\pm=(\pm\lambda_s,\vk_s,m),\quad
\gX_{sl}^\pm=(\pm\lambda_{sl},\vk_s-\vk_l,m_l).
\end{align}
In the spherical coordinates, the elementary integral is given by
\begin{align}
\label{gZdef}
\mathcal{Z}^\pm_L\left(\gX^+_1,\ldots,\gX_L^+;\gT\right)&=\frac{1}{2^L\pi}\elimit
\intE p\,f(\pm E-\mu)
\int\limits_{-1}^1\rmd x\int\limits_{0}^{2\pi}\rmd\phi\prod_{l=1}^L\frac{1}{\lambda_l(E-E_{0l}-i\epsilon)+\vc{p}\vc{k}_l}
\nl
&=\rp\mathcal{Z}^\pm_L+i\,\ip\mathcal{Z}^\pm_L,
\end{align}
where $\vc{p}=p\,\left(\cos\phi\sqrt{1-x^2},\sin\phi\sqrt{1-x^2},x\right)$,
\begin{align}
E_{0l}&\equiv E_{0}(\varepsilon_l,\lambda_l)=\frac{\varepsilon_l}{\lambda_l}\,,\quad
\varepsilon_l\equiv \varepsilon(\gX_l^+,m)=\varepsilon(\gX_l^-,m)=\frac{k_l^2+m_l^2-\lambda_l^2-m^2}{2},\quad k_l=|\vc{k}_l|,
\end{align}
and
$$
\Lambda_E=\sqrt{\Lambda^2+m^2}.
$$

So we have reduced the task to evaluating functions of the same kind. For further consideration, we need to write the components of $\vc{k}_l$:
\begin{align}
\vc{k}_l=k_l(\sin\delta_l\cos\phi_l,\sin\delta_l\sin\phi_l,\cos\delta_l).
\end{align}
Let us also introduce
\begin{align}
\label{psils}
\cos\psi_{ls}&=\frac{\vc{k}_l\vc{k}_s}{k_lk_s},\quad \sin\psi_{ls}=\abs{\frac{\vc{k}_l\times\vc{k}_s}{k_lk_s}},
\end{align}
where $\vc{a}\times\vc{b}$ and $\vc{a}\vc{b}$ denote the vector and scalar products of two 3-vectors, respectively.
As one can easily see,
\begin{align}
\label{cospsils}
\cos\psi_{ls}&=\cos\delta_l\cos\delta_s+\sin\delta_l\sin\delta_s\cos(\phi_l-\phi_s).
\end{align}
One should keep in mind that $\delta_l,\psi_{ls}\in[0,\pi]$ and so we always have $\sin\delta_l,\,\sin\psi_{ls}\geq0$. Also, without loss of generality, one can choose
\begin{align}
\label{Klevsystem}
\vk_1=k_1(0,0,1),\quad \vk_2=k_2(\sin\delta_2,0,\cos\delta_2).
\end{align}

At the end of this section, let us discuss how Eq. (\ref{gZdef}) can be converted to integrals of real functions. The method is based on applying the famous formula
\begin{align*}
\elimit\frac{1}{x-i\epsilon}=\mathcal{P}\frac1x+i\pi\delta(x)
\end{align*}
which can be generalized as
\begin{align}
\label{elimit}
\elimit\frac{1}{x-i\lambda\epsilon}=\mathcal{P}\frac1x+i\pi\,\sgn(\lambda)\,\delta(x),\quad \lambda\neq0.
\end{align}
Writing the product under the integral (\ref{gZdef}) as
\begin{align}
\label{Pdelta}
\prod_{l=1}^L\frac{1}{\lambda_l(E-E_{0l}-i\epsilon)+\vc{p}\vc{k}_l}&=
\prod_{l=1}^L\left[\Vp\frac1{\lambda_l(E-E_{0l})+\vc{p}\vc{k}_l}+i\pi\,\sgn\lambda_l\,\delta\left(\lambda_l(E-E_{0l})+\vc{p}\vc{k}_l\right)\right]
\end{align}
and expanding it, we obtain separate real and imaginary parts. Equation (\ref{Pdelta}) assumes that one takes $L$ different limits $\epsilon_l\rightarrow0$. This approach coincides with the one used in \cite{Klevansky} for calculation of the three line integral for non-zero momenta. However, when $k_1=k_2=0$, the authors apply a different expansion, taking only one limit $\epsilon\rightarrow0$. It is quite clear that the results obtained with both approaches must coincide since cannot depend on how we take a separate limit. As is shown below, this is true only if one takes into account the conditions for the existence of the appearing improper multiple integrals. To underline this conclusion, we will use Eq.~(\ref{Pdelta}) excluding the simplest case when all momenta are equal to zero. As one can easily see, in such case the product of two or more delta-functions immediately vanishes if $E_{0l}\neq E_{0s}$.

\section{Correction of the result \cite{Klevansky} for the three-line integral}

Before calculating $\mathcal{Z}^\pm_3$, we would like to make some remarks related to the results for $L=2$ obtained in~\cite{Klevansky}, see Eq.~(5.37). The authors of~\cite{Klevansky} derived the formulae
\begin{align}
\label{Klevint0}
\mathcal{P}\int\limits_{-1}^1\rmd x
\frac{\sgn(z+x\cos\psi)\,\thfunc{\Delta(x,z,\cos\psi)}}{(x+y)\sqrt{\Delta(x,z,\cos\psi)}}
&=
\frac{\Xi(y,z,\cos\psi)\,\thfunc{\Delta_0(y,z,\cos\psi)}+\mathscr{G}(y,z,\cos\psi)\,\thfunc{-\Delta_0(y,z,\cos\psi)}}
{\sqrt{|\Delta_0(y,z,\cos\psi)|}}\,,
\end{align}
where $\sin\psi>0$,
\begin{align}
\Delta(a,b,c)&=a^2+b^2+c^2+2\,abc-1
=(c+ab)^2-(1-a^2)(1-b^2),
\\
\Delta_0(a,b,c)&=\Delta(-a,b,c)=\Delta(a,-b,c),
\end{align}
and
\begin{align}
\label{Gdef}
\mathscr{G}(a,b,c)&\triangleq\sgn(b+c)\arccos \frac{(1+a)(a-bc)-\Delta_0(a,b,c)}{(1+a)\sqrt{(1-b^2)(1-c^2)}}
\nl
&\quad-\sgn(b-c0\arccos \frac{-(1-a)(a-bc)-\Delta_0(a,b,c)}{(1-a)\sqrt{(1-b^2)(1-c^2)}},\quad \Delta_0(a,b,c)<0,
\\
\label{Xidef0}
\Xi(a,b,c)
&\triangleq\sgn(b+c)F(1;a,b,c)-\sgn(b-c)F(-1;a,b,c)
\nl
&+\thfunc{1-b^2}\,\Big[\sgn(b-c)F(x^-;a,b,c)-\sgn(b+c)F(x^+;a,b,c)\Big],\quad \Delta_0(a,b,c)>0
\end{align}
with\footnote{The function $F(x;y,z,\psi)$ is defined up to an additive constant.}
\begin{align*}
F(x;a,b,c)&=\pm\ln\left|\frac{(x+a)(a-bc)-\Delta_0(a,b,c)\pm\sqrt{\Delta_0(a,b,c) \Delta(x,b,c)}}{x+a}\right|.
\end{align*}
Above $x^\pm$ are the roots of the quadratic $x$-trinomial $\Delta(x,b,c)$,
\begin{align*}
x^\pm&\equiv x^\pm(b,c)=-bc\pm \sqrt{d(b,c)},\quad d(b,c)\geq0,
\end{align*}
and
\begin{align}
d(b,c)=(1-c^2)(1-b^2)
\end{align}
is the corresponding determinant.

Since to evaluate the box integral we need Eq. (\ref{Klevint0}) or its generalization to a more complicated rational function, let us consider it in more detail. The first, cosmetic, disadvantage of Eqs. (\ref{Gdef}) and (\ref{Xidef0}) is that they are not explicitly symmetric with respect to the transposal $y\leftrightarrow z$. The mentioned symmetry follows from Eq. (\ref{gZdef}), see \cite{Klevansky} and below. The original expressions for both functions are also too complicated. Fortunately, we can simplify both functions by restoring explicit symmetry. First of all, it is more convenient to choose
\begin{align}
F(x;a,b,c)&=\frac12\,\ln\left|\frac{\Delta_0(a,b,c)-(x+a)(a-bc)-\sqrt{\Delta_0(a,b,c) \Delta(x,b,c)}}{\Delta_0(a,b,c)-(x+a)(a-bc)+\sqrt{\Delta_0(a,b,c) \Delta(x,b,c)}}\right|
\end{align}
which has a nice property $F(x^\pm;a,b,c)=0$. As a result, we see that the function $\Xi(a,b,c)$ is independent of $x^\pm$ and is completely defined by the first line of Eq. (\ref{Xidef0}).
After some transformations, one can prove
\begin{align}
\label{Xifunc}
\Xi(a,b,c)
&=\ln\abs{\frac{c-ab-\sqrt{\Delta_0(a,b,c)}}{c-ab+\sqrt{\Delta_0(a,b,c)}}}\,.
\end{align}
Further, mentioning that
$$
\sgn(b\pm c)=\sgn(b)\thfunc{b^2-c^2}\pm\sgn(c)\thfunc{c^2-b^2}
$$
and using Eqs. (\ref{arccossum}), one can show
\begin{align}
\label{Gfunc}
\mathscr{G}(a,b,c)&=
\sgn(c-ab)\arccos\frac{\abs{\Delta_0(a,b,c)}-(c-ab)^2}{(1-a^2)(1-b^2)}
=2\arctan\frac{c-ab}{\sqrt{\abs{\Delta_0(a,b,c)}}}\,.
\end{align}

The second, crucial, disadvantage of Eq. (\ref{Klevint0}) is that it contains excess terms. This statement can be demonstrated in two ways. On the one hand, the l.h.s.  of Eq. (\ref{Klevint0}) is obtained after integration over $\phi$ with the help of \cite{Klevansky}
\begin{align}
\label{int_1over_cos}
\int\limits_0^{2\pi}\,\frac{\rmd \phi}{a+b\cos\phi}&=\frac{2\pi}{\sqrt{a^2-b^2}}\,\sgn(a)\,\thfunc{a^2-b^2}
\end{align}
which can be generalized as
\begin{align}
\label{phicosint}
\int\limits_0^{2\pi}\,\frac{\rmd \phi}{a+b\sin\phi+c\cos\phi}
&=\frac{2\pi}{\sqrt{a^2-b^2-c^2}}\,\sgn(a)\,\thfunc{a^2-b^2-c^2}.
\end{align}
The proof is similar to that for Eq. (\ref{int_1over_cos}) and can be made using \cite{Gradshteyn}. As both last expressions are written, one assumes that the $\phi$-integral is zero when $b^2+c^2>a^2$. However, let us remind that generally in calculating $\rp\, \mathcal{Z}^\pm_2$, when the spherical coordinates  are chosen as (\ref{Klevsystem}), we deal with the triple improper integral of second kind
\begin{align}
\label{Z2integral}
&\int\limits_{m}^{\Lambda_E}\rmd E\int\limits_{-1}^1\rmd x\int\limits_{0}^{2\pi}\rmd\phi\,\frac{f(\pm E-\mu)}{p}\frac{1}{z_1+x}
\frac{1}{z_2+x\cos\delta_2+\cos\phi\sin\delta_2\cos\phi_2\sqrt{1-x^2}}\,,
\end{align}
where
\begin{align}
\label{zl}
z_l&=\frac{\lambda_lE-\varepsilon_l}{p\,k_l}.
\end{align}
The theorem on the unconditional convergence of an improper multiple integral of second kind states that the integral~(\ref{Z2integral}) unconditionally convergent that is equivalent to the existence only if the same is true for the integral
\begin{align}
\label{absoluteZ2}
&\Vp\int\limits_{m}^{\Lambda_E}\rmd E\int\limits_{-1}^1\rmd x\int\limits_{0}^{2\pi}\rmd\phi\,\left|\frac{f(\pm E-\mu)}{p}\frac{1}{z_1+x}
\frac{1}{z_2+x\cos\delta_2+\cos\phi\sin\delta_2\cos\phi_2\sqrt{1-x^2}}\right|
\nl
&=\Vp\int\limits_{m}^{\Lambda_E}\rmd E\frac{f(\pm E-\mu)}{p}\int\limits_{-1}^1\frac{\rmd x}{\left|z_1+x\right|}\int\limits_{0}^{2\pi}\frac{\rmd\phi}{\left|z_2+x\cos\delta_2+\cos\phi\sin\delta_2\cos\phi_2\sqrt{1-x^2}\right|}.
\end{align}
It can be easily shown that $\int\limits_0^{2\pi}\,\frac{\rmd \phi}{|a+b\sin\phi+c\cos\phi|}$ does not exist when $a^2<b^2+c^2$. When $a^2>b^2+c^2$,
\begin{align}
\int\limits_0^{2\pi}\,\frac{\rmd \phi}{|a+b\sin\phi+c\cos\phi|}&=\frac{2\pi}{\sqrt{a^2-b^2-c^2}}.
\end{align}
Continuing evaluation of Eq. (\ref{absoluteZ2}), we need to calculate
\begin{align}
\label{absoluteZ2:xint}
&\Vp\int\limits_{-1}^1\frac{\rmd x}{\left|z_1+x\right|\sqrt{\Delta_2(x)}},
\end{align}
where we introduce $\Delta_l(x)=\Delta(x,z_l,\cos\delta_l)$ and $\Delta_2(x)$ plays the role of $a^2-b^2-c^2$. It can be proved that the integral~(\ref{absoluteZ2:xint}) is finite only if $1-z_1^2<0$ and exists if $\Delta_2(x)$ is positive for any $x\in(-1,1)$ since otherwise we have a contribution $\int\limits_{x^-}^{x^+}\rmd x\ldots$, where $\sqrt{\Delta_2(x)}$ is not a real function. As a result, we have to assume $1-z_2^2<0$.

Summarising, we conclude that the correct version of Eq. (\ref{Klevint0}) is
\begin{align}
\label{Klevintcorrect}
\mathcal{P}\int\limits_{-1}^1\rmd x
\frac{\sgn(z+x\cos\psi)}{(x+y)\sqrt{\Delta(x,z,\cos\psi)}}
&=\frac{\Xi(y,z,\cos\psi)}
{\sqrt{\Delta_0(y,z,\cos\psi)}},
\end{align}
where $1-z^2<0$ and $1-y^2<0$ are assumed. We see that the region of the existence is symmetric until the transposal $y\leftrightarrow z$.

On the other hand, if we believe that Eqs. (\ref{phicosint}) and (\ref{Klevint0}) can be used without any changes, application of Eq. (\ref{Klevint0}) to calculate $\rp\, \mathcal{Z}^\pm_2$ when the spherical coordinates are chosen as (\ref{Klevsystem}) and $k_{1,2}>0$ immediately gives
\begin{align*}
\mathcal{P}\int\limits_{-1}^1\rmd x
\frac{\sgn(z_s+x\cos\delta_s)\,\thfunc{\Delta_s(x)}}{(x+z_l)\sqrt{\Delta_s(x)}}
&=
\frac{\Xi_{ls}(E)\,\thfunc{\Delta_{ls}}+\mathscr{G}_{ls}(E)\,\thfunc{-\Delta_{ls}}}{\sqrt{|\Delta_{ls}|}}\,,
\end{align*}
where we introduce $\Delta_{ls}=\Delta_0(z_l,z_s,\cos\psi_{ls})=\Delta_l(-z_s)=\Delta_s(-z_l)$, $\Xi_{ls}(E)=\Xi(z_l,z_s,\cos\psi_{ls})$, and $\mathscr{G}_{ls}(E)=\mathscr{G}(z_l,z_s,\cos\psi_{ls})$ for shortness.

However, we can use other spherical coordinates demanding that $\phi_1=\phi_2$ while $\sin\delta_{1,2}>0$. Then
\begin{align*}
&\frac1{2\pi}\,\Vp\int\limits_{-1}^1\rmd x\int\limits_{0}^{2\pi}\rmd\phi\,\frac{1}{z_1+x\cos\delta_1+\sqrt{1-x^2}\sin\delta_1(\sin\phi\sin\phi_1+\cos\phi\cos\phi_1)}
\nl
&\quad\times\frac{1}{z_2+x\cos\delta_2+\sqrt{1-x^2}\sin\delta_2(\sin\phi\sin\phi_1+\cos\phi\cos\phi_1)}
\nl
&=\sin\delta_1\int\limits_{-1}^1\rmd x
\frac{\sgn(z_1+x\cos\delta_1)\,\thfunc{\Delta_1(x)}}{(\beta_{12}+\alpha_{12}x)\sqrt{\Delta_1(x)}}
-\sin\delta_2\int\limits_{-1}^1\rmd x
\frac{\sgn(z_2+x\cos\delta_2)\,\thfunc{\Delta_2(x)}}{(\beta_{12}+\alpha_{12}x)\sqrt{\Delta_2(x)}}\,.
\end{align*}
Above we have used Eq. (\ref{phicosint}) and denoted
\begin{align}
\alpha_{ls}&=\cos\delta_s\sin\delta_l-\cos\delta_l\sin\delta_s\cos(\phi_l-\phi_s),
\\
\beta_{ls}&=z_s\sin\delta_l-z_l\sin\delta_s\cos(\phi_l-\phi_s),
\end{align}
for arbitrary spherical coordinates.

Applying Eq. (\ref{Klevint0}) to every term separately, mentioning that $\alpha_{12}^2=\sin^2\psi_{12}$,
\begin{align}
\label{Deltabetaalpha}
\Delta_t\left(-\frac{\beta_{ls}}{\alpha_{ls}}\right)=\frac{\sin^2\delta_t}{\sin^2\psi_{ls}}\,\Delta_{ls},
\quad t=l,s,
\end{align}
and using Eq. (\ref{arctansum}), we obtain
\begin{eqnarray*}
\label{piterm3line}
&\sin\delta_1&\int\limits_{-1}^1\rmd x
\frac{\sgn(z_1+x\cos\delta_1)\,\thfunc{\Delta_1(x)}}{(\beta_{12}+\alpha_{12}x)\sqrt{\Delta_1(x)}}
-\sin\delta_2\int\limits_{-1}^1\rmd x
\frac{\sgn(z_2+x\cos\delta_2)\,\thfunc{\Delta_2(x)}}{(\beta_{12}+\alpha_{12}x)\sqrt{\Delta_2(x)}}
\nl
&=&\frac{\sin\psi_{12}}{\alpha_{12}}\Bigg[\frac{\Xi(\beta_{12}/\alpha_{12},z_1,\cos\delta_1)\,\thfunc{\Delta_{12}}
+\mathscr{G}(\beta_{12}/\alpha_{12},z_1,\cos\delta_1)\,\thfunc{-\Delta_{12}}}{\sqrt{|\Delta_{12}|}}
\nl
&&-\frac{\Xi(\beta_{12}/\alpha_{12},z_2,\cos\delta_2)\,\thfunc{\Delta_{12}}
+\mathscr{G}(\beta_{12}/\alpha_{12},z_2,\cos\delta_2)\,\thfunc{-\Delta_{12}}}{\sqrt{|\Delta_{12}|}}\Bigg]
\nl
&=&\frac{\Xi_{12}(E)\,\thfunc{\Delta_{12}}+[\mathscr{G}_{12}(E)-\pi]\,\thfunc{-\Delta_{12}}}{\sqrt{|\Delta_{12}|}}
\end{eqnarray*}
which contains the additional $\pi$-term when $\Delta_{12}<0$. Since the double integral cannot depend on the coordinate system, the mentioned inconsistency proves that the integral does not exist for $\Delta_{12}<0$. Using the symmetry between $z_1$ and $z_2$, we conclude that the conditions for the existence are $1-z_{1,2}^2<0$.

Besides Eq. (\ref{Z2integral}), $\rp \mathcal{Z}_2^\pm$ contains a second term
\begin{align}
\label{Z2deltadelta}
&\int\limits_{m}^{\Lambda_E}\rmd E\int\limits_{-1}^1\rmd x\int\limits_{0}^{2\pi}\rmd\phi\,\frac{f(\pm E-\mu)}{p}\,\delta(z_1+x)\,
\delta\left(z_2+x\cos\delta_2+\cos\phi\sin\delta_2\cos\phi_2\sqrt{1-x^2}\right)
\nl
&=2\int\limits_{m}^{\Lambda_E}\rmd E\frac{f(\pm E-\mu)}{p}\frac{\Theta(-\Delta_{12})}{\sqrt{|\Delta_{12}|}}
\end{align}
obtained with the help of the well-known result
\begin{align}
\label{xint_delta}
\int\limits_{-1}^1\rmd x\,f(x)\,\delta(a+x)&=f(-a)\thfunc{1-a^2}
\end{align}
and
\begin{align}
\label{fdeltasincosphi}
&\int\limits_{0}^{2\pi}\rmd\phi\:f(\phi)\,\delta\left(a+b\cos\phi+c\sin\phi\right)=
\frac{f\left(\phi^+\right)+f\left(\phi^-\right)}{\sqrt{|a^2-b^2-c^2|}}\,\thfunc{b^2+c^2-a^2},
\end{align}
where $\phi^\pm$ are defined by the relations
\begin{align*}
\cos\phi^\pm&=\frac{-ab\pm c\,\sqrt{b^2+c^2-a^2}}{b^2+c^2},\quad\sin\phi^\pm=\frac{-ac\mp b\,\sqrt{b^2+c^2-a^2}}{b^2+c^2}.
\end{align*}
The integral (\ref{Z2deltadelta}) evidently exists and is non-zero only when $\Delta_{12}<0$. We see that the integration of a $\delta$-function does not influence the existence of a multiple integral and just leads to $\Theta$-expressions. More precisely saying, the conditions for the existence  must be investigated after making all integrations of $\delta$-functions. If the resulting expression is an ordinary integral, it exactly exists if it is finite.

Combining two considered contributions, we find that the second term vanishes,
\begin{align}
\rp \mathcal{Z}_2^\pm
&=
\frac{1}{2}
\intE\,\frac{f(\pm E-\mu)\,\Xi_{12}(E)}{\sqrt{\Phi_{12}(E)}},
\end{align}
and the conditions $\Phi_l(E)\leq0$, $l=1,2$ must be satisfied. The quadratic trinomials $\Phi_l=p^2k_l^2(1-z_l^2)$ and $\Phi_{ls}=p^2k_l^2k_s^2\Delta_{ls}$ are described in detail in Appendix B.

A key difference of the condition for the existence from a simple cut off by the $\Theta$-function is that the former must be valid for any $E\in[m, \Lambda_E]$ that entails the necessity of finite $\Lambda_E$ for evaluation of integrals with three and greater fermion lines. For $\rp \mathcal{Z}_2^\pm$, Eq.~(\ref{posOmega})
results in the necessary constraints $k_l\leq|\lambda_l|$.

It can be shown in a similar way that the factor $\Theta(\Delta_{12})=\Theta(\Phi_{12})$ in the imaginary part $\ip \mathcal{Z}_2^\pm$ that can be extracted from Eq.~(5.44) in \cite{Klevansky}, should be considered as the condition for the existence.


\section{The $Z_3^*$ for collinear momenta}
Let us now start the detailed evaluation of $\mathcal{Z}^\pm_3$ for all possible orientations of momenta $\vc{k}_{1,2,3}$. As one can easily check, up to permutation of $\vc{k}_{1,2,3}$ we have to consider seven opportunities:
\begin{enumerate}
  \item $k_1=k_2=k_3=0$;
  \item $k_1=k_2=0$, $k_3>0$;
  \item $k_l>0$, $\vecprod{\vk_l}{\vk_s}=0$;
  \item $k_{1,2}>0$, $k_3=0$, $\vc{k}_1\times\vc{k}_2=0$;
  \item $k_{1,2}>0$, $k_3=0$, $\vc{k}_1\times\vc{k}_2\neq0$;
  \item $k_l>0$, $\vc{k}_l\times\vc{k}_s\neq0$, $\vc{k}_1(\vc{k}_2\times\vc{k}_3)=0$;
  \item $k_l>0$, $\vc{k}_l\times\vc{k}_s\neq0$, $\vc{k}_1(\vc{k}_2\times\vc{k}_3)\neq0$.
\end{enumerate}
The first four cases corresponding to {\it``dot''} and {\it ``line''} configurations are very simple. For this reason, it is more convenient to derive a general expression for $\gZ_L^\pm$ and then to apply it to $L=3$. These cases will be considered below in this section. Planar configurations (items 5 and 6) when all three vectors $\vc{k}_1$, $\vc{k}_2$, and $\vc{k}_3$ lie in the same plane or, more precisely, when $\vc{k}_1(\vc{k}_2\times\vc{k}_3)=0$ are considered in the next section. The most complicated situation for three independent vectors is analyzed separately in Sec. VI.

\subsection{Calculation of $\gZ_L^\pm$ when all $k_l=0$}
The simplest situation is when all $k_l$ are equal to zero. Equation (\ref{gZdef}) is strongly simplified and takes the form
\begin{align}
\label{gZk0}
\mathcal{Z}^\pm_L(\gX_{1}^+,\ldots,\gX_{L}^+;\gT)
&=
\frac{1}{2^{L-2}\rho_L}\elimit
\intE p\,f(\pm E-\mu;T)\prod_{l=1}^L\frac{1}{E-E_{0l}-i\epsilon}
\end{align}
with
\begin{align}
\rho_L&=\prod_{l=1}^L\lambda_l.
\end{align}

To calculate the integral, we need to translate the product into a sum. Let us here remind the useful relation \cite{Gradshteyn}
\begin{align}
\frac{\varphi(x)}{f(x)}&=\sum_{l=1}^L\frac{c'_l}{x-x_l},\quad c'_l=\frac{\varphi(x_l)}{f'(x_l)},
\end{align}
where we assume that the function $f(x)$ has only simple roots $x_l$. A consequence of this wonderful expression is the relation
\begin{align}
\label{decomposition}
\prod_{l=1}^L\frac{1}{\alpha_l y-y_l}&=\sum_{l=1}^L\frac{c_l^{(L)}}{\alpha_l y-y_l},
\end{align}
where $y_s\neq y_l$ if $s\neq l$, and
\begin{align}
c_1^{(1)}=1,\quad c_l^{(L)}=\prod_{s=1,s\neq l}^L\frac{\alpha_l}{y_l\alpha_s-y_s\alpha_l},\quad L>1.
\end{align}

Using Eq. (\ref{decomposition}), the integral (\ref{gZk0}) can be written as a sum of simple integrals, and one obtains
\begin{align}
\label{ReZK0}
\rp\mathcal{Z}^\pm_L
&=
\frac{1}{2^{L-2}\rho_L}\sum_{l=1}^La_l^{(L)}\,I_G^\pm(E_{0l};\gT),\\
\label{ImZK0}
\ip\mathcal{Z}^\pm_L
&=
\frac{\pi}{2^{L-2}\rho_L}\sum_{l=1}^La_l^{(L)}\,G^\pm(E_{0l};\gT).
\end{align}
The assumption is used that for every pair $E_{0s}\neq E_{0l}$ if $s\neq l$. Here
\begin{align}
a_1^{(1)}&=1,\quad a_l^{(L)}
=
\prod_{s=1,s\neq l}^L\frac1{E_{0l}-E_{0s}},\quad L>1,
\end{align}
\begin{align}
\label{IGdef}
I_G^\pm(E_0;\gT)&=\Vp\intE\,\frac{pf(\pm E-\mu;T)}{E-E_0},
\end{align}
and
\begin{align}
\label{FGfunc}
G^\pm(E;\gT)&=p\,f(\pm E-\mu;T)\,\thfunc{E-m}\,\thfunc{\Lambda_E-E}.
\end{align}
Considering the corresponding integral of absolute value and separating the singularity, one can show that we must demand $E_{0l}<m$ or $E_{0l}>\Lambda_E$ to have a real part while the imaginary one exists for any $E_{0l}$.

For $L=3$, one can find
\begin{align*}
a_1^{(3)}
&=\frac1{(E_{02}-E_{01})(E_{03}-E_{01})},\quad
a_2^{(3)}
=-\frac1{(E_{02}-E_{01})(E_{03}-E_{02})},\quad
a_3^{(3)}
=\frac1{(E_{03}-E_{01})(E_{03}-E_{02})}
\end{align*}
which are connected by
\begin{align*}
  a_1^{(3)}+a_2^{(3)}+a_3^{(3)}=0.
\end{align*}

\subsection{Calculation of $\gZ_L^\pm$ for $\vecprod{\vk_l}{\vk_s}=0$}
The condition $\vecprod{\vk_l}{\vk_s}=0$ for some $l,s$ means that $\vc{k}_l$ and $\vc{k}_s$ are collinear if $k_lk_s\neq0$. In this case,
using the coordinates~(\ref{Klevsystem}), we have
\begin{align}
\label{etals}
\cos\psi_{ls}&\equiv\eta_{ls}=\pm1,\quad\sin\psi_{ls}=0,
\\
\sin\delta_l&=0,\quad \cos\delta_l=\eta_{ls}\cos\delta_s=\pm1.\nonumber
\end{align}
Another possibility is that $k_l=0$ or/and $k_s=0$.

Below we assume $L>1$ since we need two or more vectors to speak about collinearity\footnote{The solution for the case $L=1$ can be found in \cite{Klevansky}. We only mention that values of the momentum $k_1$ for which  $\rp \mathcal{Z}^\pm_1$ exists, are determined from the inequality $\Phi_1(E)\leq0$.}. Equation (12) can be written as
\begin{align*}
\mathcal{Z}^\pm_L(\gX_{1}^+,\ldots,\gX_{L}^+;\gT)
&=
\frac{1}{2^{L}\pi}\elimit
\intE p\,f(\pm E-\mu;T)\prod_{s=L-n+1}^L
\frac{1}{\lambda_s(E-E_{0s}-i\epsilon)}
\nl
&\times\int\limits_{-1}^1\rmd x\int\limits_0^{2\pi}\rmd\phi\prod_{l=1}^{L-n}
\frac{1}{\lambda_l(E-E_{0l}-i\epsilon)+xpk_{l}\cos\delta_l }
\end{align*}
where $n=\overline{0,L-1}$. If $n=0$, the first ($E$-) product is absent.

Let us define
\begin{align}
\label{omegazetadef}
\omega_{ls}&=k_l\lambda_s-k_s\lambda_l\eta_{ls},\quad
\zeta_{ls}=\frac{k_l\varepsilon_s-k_s\varepsilon_l\eta_{ls}}{\omega_{ls}}.
\end{align}
One can see that if $k_s=0$, $\zeta_{ls}=E_{0s}$. When $k_lk_s\neq0$, due to $\eta_{ls}^2=1$, we have two nice properties:
\begin{align}
\omega_{ls}&=-\omega_{sl}\eta_{ls},\quad \zeta_{ls}=\zeta_{sl}.
\end{align}

Collinearity of $\vc{k}_l$ allows us to decompose the product over $x$, according to Eq. (\ref{decomposition}), as
\begin{align}
\label{collinearproduct}
&\prod_{l=1}^{L'}
\frac{1}{\lambda_l(E-E_{0l}-i\epsilon)+xpk_{l}\cos\delta_l}
=\sum_{l=1}^{L'}
\frac{1}{\lambda_l(E-E_{0l}-i\epsilon)+xpk_{l}\cos\delta_l}
\prod_{s=1,s\neq l}^{L'}\frac{k_l}{(E-\zeta_{ls}-i\epsilon)\,\omega_{ls}}.
\end{align}
Then we again apply Eq. (\ref{decomposition}) to the two-component $E$-product
\begin{equation}
\label{Eproduct}
\prod_{s=1,s\neq l}^{L'}\frac{k_l}{(E-\zeta_{ls}-i\epsilon)\,\omega_{ls}}\prod_{s=L'+1}^{L''}\frac{1}{\lambda_s(E-\zeta_{ls}-i\epsilon)}=k_l\sum_{s=1,s\neq l}^{L''}\frac{b\find{L''}_{ls}}{E-\zeta_{ls}-i\epsilon},
\end{equation}
where
\begin{align*}
b\find{L''}_{ls}&=\frac{d\find{L''}_{ls}}{\omega_{ls}},\quad
d\find{2}_{12}=1,\quad d\find{L''}_{ls}=\prod_{t=1,t\neq s,t\neq l}^{L''}\frac{k_l}{\omega_{lt}(\zeta_{ls}-\zeta_{lt})}.
\end{align*}
and, by definition, $L'<L''$. These coefficients have the following  nice properties:
\begin{align}
b\find{L''}_{ls}&=-b\find{L''}_{sl}\eta_{ls},\quad d\find{L''}_{ls}=d\find{L''}_{sl}.
\end{align}
Combining two above decompositions and using Eq. (\ref{Pdelta}), one gets
\begin{align}
&p\prod_{s=L-n+1}^{L}\frac{1}{\lambda_s(E-E_{0s}-i\epsilon)}\prod_{l=1}^{L-n}
\frac{1}{\lambda_l(E-E_{0l}-i\epsilon)+xpk_{l}\cos\delta_l}
\nl
&=\sum_{l=1}^{L-n}
\frac{pk_l}{\lambda_l(E-E_{0l}-i\epsilon)+xpk_{l}\cos\delta_l}
\sum_{s=1,s\neq l}^L\frac{b\find{L}_{ls}}{E-\zeta_{ls}-i\epsilon}.
\end{align}
The $\phi$-integration is trivial. Separating the real and imaginary parts, we obtain
\begin{align}
\rp\mathcal{Z}^\pm_L&=\frac{1}{2^{L-1}}\sum_{l=1}^{L-n}\sum_{s=1,s\neq l}^Lb_{ls}^{(L)}
\intE f(\pm E-\mu)\int\limits_{-1}^1\rmd x
\left[\frac{1}{z_l+x\cos\delta_l}\frac{1}{E-\zeta_{ls}}
-\pi^2\sgn\lambda_l\,\delta(E-\zeta_{ls})\,\delta(z_l+x\cos\delta_l)\right],
\\
\ip\mathcal{Z}^\pm_L&=\frac{\pi}{2^{L-1}}\sum_{l=1}^{L-n}\sum_{s=1,s\neq l}^Lb_{ls}^{(L)}
\intE f(\pm E-\mu)\int\limits_{-1}^1\rmd x\left[\frac{\delta(E-\zeta_{ls})}{z_l+x\cos\delta_l}
+\frac{\sgn\lambda_l}{E-\zeta_{ls}}\,\delta(z_l+x\cos\delta_l)\right].
\end{align}

To define constraints on momenta, let us consider the integrals of the corresponding amplitudes. One can find that the first term of the real part exists if we have $1-z_l^2\leq0$ for $E\in[m,\Lambda_E]$ and $\xi_{ls}\not\in[m,\Lambda_E]$. As a consequence, the second term is equal to zero:
\begin{align}
\label{ReZparallel_prelim}
\rp\mathcal{Z}^\pm_L&=
\frac{1}{2^{L-1}}\sum_{l=1}^{L-n}\sum_{s=1,s\neq l}^Lb_{ls}^{(L)}I_J^\pm(\zeta_{ls},k_l,\lambda_l,\varepsilon_l,\gT)
\end{align}
with
\begin{align}
I_J^\pm(E_0,k,\lambda,\varepsilon,\gT)=\intE\frac{f(\pm E-\mu)}{E-E_0}\ln\left|\frac{\lambda E-\varepsilon+ k\,p}{\lambda E-\varepsilon-k\,p}\right|.
\end{align}
As was mentioned above, the inequality $1-z_l^2\leq0$ can be satisfied only when $k_l\leq|\lambda_l|$.

The second term of the imaginary part exists when $\xi_{ls}\not\in[m,\Lambda_E]$. Then the first term is zero if $|z_l(\xi_{ls})|>1$ or, equivalently, $\Phi_l(\xi_{ls})\leq0$. Applying Eq. (\ref{xint_delta}), we finally have
\begin{align}
\label{ImZparallel_prelim}
\ip\mathcal{Z}^\pm_L&=\frac{\pi}{2^{L-1}}\sum_{l=1}^{L-n}\sgn\lambda_l\sum_{s=1,s\neq l}^Lb_{ls}^{(L)}I_H^\pm(\zeta_{ls},k_l,\lambda_l,\varepsilon_l;\gT),
\\
I_H^\pm(E_0,k_l,\lambda_l,\varepsilon_l;\gT)&=\int\limits_{m}^{\Lambda_E}\rmd E\,\frac{f(\pm E-\mu)}{E-E_0}\,\thfunc{\Phi_l\left(E\right)}.
\end{align}

Equations (\ref{ReZparallel_prelim}) and (\ref{ImZparallel_prelim}), in principle, are sufficient for numerical calculations, but it is useful to make some additional transformations which allow one to clarify the structure of the expressions. Using $\zeta_{ls}=\zeta_{sl}$, we can make resummation of the symmetric terms. Separating the part with $s>L-n$, one obtains
\begin{align}
\rp\mathcal{Z}^\pm_L&=
\frac{1}{2^{L-1}}\sum_{l=1}^{L-n-1}\sum_{s=l+1}^{L-n}b_{ls}^{(L)}\intE
\frac{f(\pm E-\mu)}{E-\zeta_{ls}}\ln\left|\frac{\lambda_lE-\varepsilon_l+k_l p}{\lambda_lE-\varepsilon_l-k_l p}\frac{\lambda_sE-\varepsilon_s-k_s\eta_{ls} p}{\lambda_sE-\varepsilon_s+k_s\eta_{ls} p}\right|\nl
&+\frac{1}{2^{L-1}}\sum_{s=L-n+1}^{L}\sum_{l=1}^{L-n}b_{ls}^{(L)}I_J^\pm(E_{0s},k_l,\lambda_l,\varepsilon_l,\gT),
\\
\ip\mathcal{Z}^\pm_L&=\frac{\pi}{2^{L-1}}\sum_{l=1}^{L-n-1}\sum_{s=l+1}^{L-n}b_{ls}^{(L)}\left[\sgn\lambda_sI_H^\pm(\zeta_{ls},k_s,\lambda_s,\varepsilon_s;\gT)
+\sgn\lambda_lI_H^\pm(\zeta_{ls},k_l,\lambda_l,\varepsilon_l;\gT)\right]
\nl
&+\frac{\pi}{2^{L-1}}\sum_{s=L-n+1}^L\sum_{l=1}^{L-n}\sgn\lambda_lb_{ls}^{(L)}I_H^\pm(E_{0s},k_l,\lambda_l,\varepsilon_l;\gT).
\end{align}


\section{Planar momenta}
\subsection{The configuration with $k_{1,2}>0$, $k_3=0$, $\vc{k}_1\times\vc{k}_2\neq0$}
Below we will work in the coordinates (\ref{Klevsystem}). As a result, the condition of noncollinearity $\vc{k}_1\times\vc{k}_2\neq0$ is equivalent to
\begin{align}
\sin\psi_{12}=\sin\delta_2>0
\end{align}
and we have
\begin{align*}
\mathcal{Z}^\pm_3(\gX_1^+,\gX_2^+,\gX_3^+;\gT)
&=
\frac{1}{8\pi\lambda_3}\elimit
\intE \frac{pf(\pm E-\mu)}{E-E_{03}-i\epsilon}
\int\limits_{-1}^1\frac{\rmd x}{\lambda_1(E-E_{01}-i\epsilon)+pk_{1}\,x}
\nl
&\quad\quad\times\int\limits_0^{2\pi}
\frac{\rmd\phi}{\lambda_2(E-E_{02}-i\epsilon)+pk_2(\sin\delta_2\cos\phi\sqrt{1-x^2}+x\cos\delta_2 )}\,.
\end{align*}

The $\phi$-integration is made using Eq. (\ref{phicosint}), if the condition $a^2>b^2+c^2$ is assumed, and Eq. (\ref{fdeltasincosphi}); meanwhile, integrals over $x$ can be evaluated with the help of Eqs. (\ref{Klevintcorrect}), (\ref{xint_delta}), and
\begin{align}
\label{xint_xsqrtR_m}
\int\limits_{-1}^{1}\rmd x\frac{\thfunc{-\Delta(x,z,\cos\psi)}}{(x+y)\sqrt{-\Delta(x,z,\cos\psi)}}
&=
\pi\,\frac{\thfunc{1-z^2}}{\sqrt{\Delta_0(y,z,\cos\psi)}}\,\sgn(y-z\cos\psi),\quad \Delta_0(y,z,\cos\psi)>0,
\end{align}
which was proved in \cite{Klevansky} excluding that positivity of $\Delta_0$ is the condition for the existence.

Finally, we find
\begin{align}
\label{ReZ3planark3}
\rp\mathcal{Z}^\pm_3&=\frac{1}{4\lambda_3}\intE\frac{f(\pm E-\mu)}{E-E_{03}}\frac{\Xi_{12}(E)}{\sqrt{\Phi_{12}(E)}},\quad \Phi_{1,2}(E)\leq0,
\\
\label{ImZ3planark3}
\ip\mathcal{Z}^\pm_3
&=\frac{\pi}{4\lambda_3}\intE \frac{f(\pm E-\mu)}{E-E_{03}}\frac{\Pi_{12}(E)}{\sqrt{\Phi_{12}(E)}},\quad \Phi_{12}(E)>0,\  \Phi_{1,2}(E_{03})<0,
\end{align}
with $E_{03}\not\in[m,\Lambda_E]$ assumed in both expressions and defined for shortness
\begin{align}
\label{Pidef}
\Pi_{ls}(E)&=\sgn\lambda_l\,\sgn(z_s-z_l\cos\psi_{ls})\,\thfunc{\Phi_l(E)}
+\sgn\lambda_s\,\sgn(z_l-z_s\cos\psi_{ls})\,\thfunc{\Phi_s(E)}
\nl
&=\sgn\left[\lambda_l\omega_{ls}(E-\zeta_{ls})\right]\,\thfunc{\Phi_l(E)}
+\sgn\left[\lambda_s\omega_{sl}(E-\zeta_{ls})\right]\,\thfunc{\Phi_s(E)}.
\end{align}

\subsection{The case when $k_l>0$, $\vc{k}_l\times\vc{k}_s\neq0$, $\vc{k}_1(\vc{k}_2\times\vc{k}_3)=0$}
First of all, let us redefine this configuration into a more appropriate form. Three noncollinearity requirements give
\begin{align}
\sin\delta_2,\ \sin\delta_3,\ \sin\psi_{23}>0;
\end{align}
meanwhile, the complanarity condition means that $\sin\phi_3=0$ and
\begin{align}
\cos\phi_3\equiv\eta_{23}=\pm1.
\end{align}
As a result, we can write
\begin{align}
\cos\psi_{23}&=\cos(\delta_2-\eta_{23}\delta_3),\quad
\alpha_{23}=\sin(\delta_2-\eta_{23}\delta_3),\quad\alpha_{32}=-\eta_{23}\alpha_{23},\quad \alpha_{23}^2=\alpha_{32}^2=\sin^2\psi_{23}
\\
\beta_{ls}&=z_s\sin\delta_l-\eta_{23}z_l\sin\delta_s,\quad \beta_{32}=-\eta_{23}\beta_{23}.
\end{align}

Let us define the quadratic trinomial
\begin{align}
\Psi_{ls}(x)&\equiv(\beta_{ls}+\alpha_{ls}x)^2+\Delta_l(x)\sin^2\delta_s\sin^2(\phi_l-\phi_s)
\end{align}
which enters into the picture for three nonzero momenta in the same way as $\Delta_l(x)$ for two ones. The symmetry $\Psi_{ls}(x)=\Psi_{sl}(x)$ can be easily checked. We also introduce\footnote{Compare with the definition of $\Delta_0$.} $\Delta_{123}=\Psi_{23}(-z_1)
$.
As one can see, the trinomial $\Phi_{123}(E)=p^2k_1^2k_2^2k_3^2\Delta_{123}$ is simplified in the considered case and can be written as
$$
\Phi_{123}(E)=\left[\vc{b}\find{3}E-\vc{a}\find{3}\right]^2,
$$
see Appendix B for details. It is useful to remain $\Phi_{123}(E)$ in the denominators of integrands in this quadratic form to make the connection with the 3D case more transparent.

Evaluation of $\mathcal{Z}^\pm_3$ for the considered case is very similar to the calculations of the three line integral in Sec. III. Equations (\ref{phicosint}) and (\ref{fdeltasincosphi}) are used to perform $\phi$-integration, Eqs. (\ref{Klevintcorrect}), (\ref{xint_delta}), and (\ref{xint_xsqrtR_m}) are sufficient for taking integrals over $x$.

To simplify the result, we need Eq. (\ref{Deltabetaalpha}) and the following property:
\begin{align*}
\eta_{23}\Xi\left(\frac{\beta_{32}}{\alpha_{32}},z_3,\delta_3\right)&
-\Xi\left(\frac{\beta_{23}}{\alpha_{23}},z_2,\delta_2\right)=-\frac{\alpha_{23}}{\sin\psi_{23}}\,\Xi_{23}.
\end{align*}
The result can be represented as a sum of functions
\begin{align}
I_{K;ls}\find{3}&=\intE f(\pm E-\mu)
\frac{(\vecprod{\vc{k}_l}{\vc{k}_s})[\vc{b}\find{3}E-\vc{a}\find{3}]}{\Phi_{123}(E)}
\frac{\Xi_{ls}(E)}{\sqrt{\Phi_{ls}(E)}},
\\
I_{L;ls}\find{3}&=\intE f(\pm E-\mu)
\frac{(\vecprod{\vc{k}_l}{\vc{k}_s})\left[\vc{b}\find{3}E-\vc{a}\find{3}\right]}{\Phi_{123}(E)}\frac{\Pi_{ls}(E)}{\sqrt{\Phi_{ls}(E)}},
\end{align}
in the form
\begin{align}
\label{ReZ3planar}
\rp\mathcal{Z}^\pm_3&=\frac{1}{4 }\left[I_{K;12}\find{3}+I_{K;23}\find{3}+I_{K;31}\find{3}\right],\quad \Phi_l(E)\leq0,
\\
\label{ImZ3planar}
\ip\mathcal{Z}^\pm_3
&=\frac{\pi}{4 }\left[I_{L;12}\find{3}+I_{L;23}\find{3}+I_{L;31}\find{3}\right],\quad \Phi_{ls}(E)\geq0,
\end{align}
where the root of $\Phi_{123}(E)$, $E_{123}=\vc{a}\find{3}\vc{b}\find{3}/[\vc{b}\find{3}]^2$, must lie outside the interval $[m,\Lambda_E]$.

\section{Box integral for three independent momenta}
\subsection{$\phi$-integration}
\label{L3product}
To calculate the box elementary function, it is needed to expand the product part of the integrand that can be explicitly written as
\begin{eqnarray}
&&\frac{pk_1}{\lambda_1(E-E_{01}-i\epsilon)+pk_1x}
\frac{pk_2}{\lambda_2(E-E_{02}-i\epsilon)+pk_2(\cos\phi\sin\delta_2\sqrt{1-x^2}+x\cos\delta_2)}
\nl
&&\quad\times\frac{pk_3}{\lambda_3(E-E_{03}-i\epsilon)+pk_3\big[(\cos\phi_3\cos\phi+\sin\phi_3\sin\phi)\sin\delta_3\sqrt{1-x^2}+x\cos\delta_3\big]}.
\end{eqnarray}
Using Eq. (\ref{Pdelta}), one finds that the real part of the product is
\begin{eqnarray}
&&\frac{1}{(z_1+x)\left(z_2+x\cos\delta_2+\cos\phi\sin\delta_2\sqrt{1-x^2}\right)}
\frac{1}{z_3+x\cos\delta_3+\cos\phi\sin\delta_3\cos\phi_3\sqrt{1-x^2}+\sin\phi\sin\delta_3\sin\phi_3\sqrt{1-x^2}}
\nl
&&-\pi^2\,\sgn(\lambda_1\lambda_2)\,\frac{\delta\left(z_1+x\right)\,\delta\left(z_2+x\cos\delta_2+\cos\phi\sin\delta_2\sqrt{1-x^2}\right)}
{z_3+x\cos\delta_3+\cos\phi\sin\delta_3\cos\phi_3\sqrt{1-x^2}+\sin\phi\sin\delta_3\sin\phi_3\sqrt{1-x^2}}
\nl
&&-\pi^2\,\sgn(\lambda_1\lambda_3)
\frac{\delta\left(z_1+x\right)\,\delta\left(z_3+x\cos\delta_3+\cos\phi\sin\delta_3\cos\phi_3\sqrt{1-x^2}+\sin\phi\sin\delta_3\sin\phi_3\sqrt{1-x^2}\right)}
{z_2+x\cos\delta_2+\cos\phi\sin\delta_2\sqrt{1-x^2}}
\nl
&&-\pi^2\,\sgn(\lambda_2\lambda_3)\frac{\delta\left(z_2+x\cos\delta_2+\cos\phi\sin\delta_2\sqrt{1-x^2}\right)}{z_1+x}
\nl
&&\quad\times\delta\left(z_3+x\cos\delta_3+\cos\phi\sin\delta_3\cos\phi_3\sqrt{1-x^2}+\sin\phi\sin\delta_3\sin\phi_3\sqrt{1-x^2}\right)
\end{eqnarray}
and the sum
\begin{eqnarray}
&&\pi\,\sgn\lambda_1\,\frac{\delta\left(z_1+x\right)}{z_2+x\cos\delta_2+\cos\phi\sin\delta_2\sqrt{1-x^2}}
\nl
&&\quad\times\frac{1}{z_3+x\cos\delta_3+\cos\phi\sin\delta_3\cos\phi_3\sqrt{1-x^2}+\sin\phi\sin\delta_3\sin\phi_3\sqrt{1-x^2}}
\nl
&&+\pi\,\sgn\lambda_2\frac{\delta\left(z_2+x\cos\delta_2+\cos\phi\sin\delta_2\sqrt{1-x^2}\right)}
{(z_1+x)\left(z_3+x\cos\delta_3+\cos\phi\sin\delta_3\cos\phi_3\sqrt{1-x^2}+\sin\phi\sin\delta_3\sin\phi_3\sqrt{1-x^2}\right)}
\nl
&&+\pi\,\sgn\lambda_3\frac{\delta\left(z_3+x\cos\delta_3+\cos\phi\sin\delta_3\cos\phi_3\sqrt{1-x^2}+\sin\phi\sin\delta_3\sin\phi_3\sqrt{1-x^2}\right)}
{(z_1+x)\left(z_2+x\cos\delta_2+\cos\phi\sin\delta_2\sqrt{1-x^2}\right)}
\nl
&&-\pi^3\,\sgn(\lambda_1\lambda_2\lambda_3)\,\delta\left(z_1+x\right)\,\delta\left(z_2+x\cos\delta_2+\cos\phi\sin\delta_2\sqrt{1-x^2}\right)
\nl
&&\quad\times\delta\left(z_3+x\cos\delta_3+\cos\phi\sin\delta_3\cos\phi_3\sqrt{1-x^2}+\sin\phi\sin\delta_3\sin\phi_3\sqrt{1-x^2}\right)
\end{eqnarray}
determines the imaginary one.

As usual, we start by integrating over $\phi$ of the product under the integral. To do this, we need the following expression:
\begin{align}
\label{doublephiint}
&\int\limits_0^{2\pi}\frac{\rmd \phi}{(a_1+b_1\cos\phi+c_1\sin \phi)(a_2+b_2\cos \phi+c_2\sin \phi)}
=
\frac{2\pi}{(a_1b_2-a_2b_1)^2-(b_1c_2-b_2c_1)^2+(a_2c_1-a_1c_2)^2}
\nl
&\times\left[\frac{a_2(b_1^2+c_1^2)-a_1(b_1b_2+c_1c_2)}{\sqrt{|a_1^2-b_1^2-c_1^2|}}\,\sgn(a_1)\,\thfunc{a_1^2-b_1^2-c_1^2}
+\frac{ a_1(b_2^2+c_2^2)-a_2(b_1b_2+c_1c_2)}{\sqrt{|a_2^2-b_2^2-c_2^2|}}\,\sgn(a_2)\,\thfunc{a_2^2-b_2^2-c_2^2}\right]
\end{align}
which is absolutely convergent when $a_1^2-b_1^2-c_1^2>0$ and $a_2^2-b_2^2-c_2^2>0$. As one can see, Eq. (\ref{phicosint}) is the limit of Eq. (\ref{doublephiint}).

Equations (\ref{fdeltasincosphi}) and (\ref{doublephiint}) allow us to make integration over $\phi$. To simplify the result, we also use the relation
\begin{align}
\delta\left([g(x)]^2-[h(x)]^2\right)&=\frac{\delta(g(x)-h(x))+\delta(g(x)+h(x))}{2\abs{h(x)}}
\end{align}
together with Eq. (\ref{fdeltasincosphi}) and find
\begin{align}
\label{deltaphiz2z3}
&\int\limits_0^{2\pi}\rmd\phi\,\delta\left(z_2+x\cos\delta_2+\cos\phi\sin\delta_2\sqrt{1-x^2}\right)
\nl
&\quad\times\delta\left(z_3+x\cos\delta_3+\cos\phi\sin\delta_3\cos\phi_3\sqrt{1-x^2}+\sin\phi\sin\delta_3\sin\phi_3\sqrt{1-x^2}\right)
\nl
&=2\sin\delta_2\sin\delta_3\abs{\sin\phi_3}\delta\left(\Psi_{23}(x)\right).
\end{align}

Applying above mentioned formulas, we obtain
\begin{align}
\label{reboxintphi}
\rp\mathcal{Z}^\pm_3&=\frac{1}{4 k_1k_2k_3}
\intE \frac{f(\pm E-\mu)}{p^2}\int\limits_{-1}^1\rmd x
\nl
&\quad\times\Bigg\{
\sin\delta_2\frac{\beta_{23}+x\alpha_{23}}{(z_1+x)\Psi_{23}(x)\sqrt{\Delta_2(x)}}\,\sgn(z_2+x\cos\delta_2)
\nl
&\quad+\sin\delta_3\frac{\beta_{32}+x\alpha_{32}}{(z_1+x)\Psi_{23}(x)\sqrt{\Delta_3(x)}}\,\sgn(z_3+x\cos\delta_3)
\nl
&\quad-\pi\,\sgn(\lambda_2\lambda_3)
\frac{\sin\delta_2\sin\delta_3\abs{\sin\phi_3}\delta\left(\Psi_{23}(x)\right)}{z_1+x}
\Bigg\},\quad 1-z_2^2\leq0,\ 1-z_3^2\leq0,
\end{align}
where the terms with $\sgn(\lambda_1\lambda_2)$ and $\sgn(\lambda_1\lambda_3)$ vanish due-to the conditions for the existence, and
\begin{align}
\ip\mathcal{Z}^\pm_3&=\frac{1}{4 k_1k_2k_3}
\intE \frac{f(\pm E-\mu)}{p^2}\int\limits_{-1}^1\rmd x
\nl
&\times\Bigg\{\sgn\lambda_2\,\sin\delta_2
\frac{\beta_{23}+x\alpha_{23}}{(z_1+x)\Psi_{23}(x)\sqrt{-\Delta_2(x)}}\,\thfunc{-\Delta_2(x)}
\nl
&\quad+\sgn\lambda_3\,\sin\delta_3\frac{\beta_{32}+x\alpha_{32}}
{(z_1+x)\Psi_{23}(x)\sqrt{-\Delta_3(x)}}\,\thfunc{-\Delta_3(x)}
\nl
&+\pi\,\sgn\lambda_1
\sin\delta_2\,\delta\left(z_1+x\right)\frac{\beta_{23}+x\alpha_{23}}{\Psi_{23}(x)\sqrt{\Delta_2(x)}}\,\sgn(z_2+x\cos\delta_2)
\nl
&\quad+\pi\,\sgn\lambda_1\sin\delta_3\,\delta\left(z_1+x\right)\frac{\beta_{32}+x\alpha_{32}}{\Psi_{23}(x)\sqrt{\Delta_3(x)}}\,
\sgn(z_3+x\cos\delta_3)
\nl
&\quad-\pi^2\,\sgn(\lambda_1\lambda_2\lambda_3)\,\delta\left(z_1+x\right)\,\sin\delta_2\sin\delta_3\abs{\sin\phi_3}\delta\left(\Psi_{23}(x)\right)
\Bigg\},\quad \Delta_{12}\geq0,\quad \Delta_{13}\geq0.
\end{align}
The terms proportional to $\sgn\,\lambda_1$ determine the conditions where the imaginary part is finite and real. Applying Eq.~(\ref{doublephiint}) leads to appearing $\Delta_{2,3}(x)$, which must be positive only at $x=-z_1$ because of the presence of $\delta(z_1+x)$.

\subsection{Two new definite integrals}
To calculate the box integral, we need to evaluate the integral
\begin{align}
\int\limits_{-1}^1\rmd x\,\frac{K\,x+S}{R_1(x)\sqrt{\Delta(x,z,\cos\delta)}}\,\sgn(z+x\cos\delta),
\end{align}
which is a generalization of Eq. (\ref{Klevintcorrect}) and the integral
\begin{align}
\int\limits_{-1}^1\rmd x\,\frac{K\,x+S}{R_1(x)\sqrt{-\Delta(x,z,\cos\delta)}}\,\thfunc{-\Delta(x,z,\cos\delta)}
\end{align}
which is an analog of Eq. (\ref{xint_xsqrtR_m}), where
\begin{align}
R_1(x)&=Cx^2+2Bx+A.
\end{align}
Both integrals can be evaluated analytically, e.g., using \cite{Gradshteyn}.

Let us start with the consideration of the first one. In the general case, integration is quite difficult. Fortunately, we can consider only a much simpler case when trinomials $\Delta(x,z,\cos\delta)$ and $R_1(x)$ are connected by the relation
\begin{align}
D&=B^2-AC=-\frac{k}{C-k}\,D_0,\quad C>k\geq0,
\end{align}
with
\begin{align}
\label{udef}
D_0=u^2-C(C-k)\,d(z,\cos\delta),\quad u=B-Cb, \quad  b=z\cos\delta.
\end{align}
It means that
\begin{align}
\label{R1Deltalink}
R_1(x)=(C-k)\left(x+b+\frac{u}{C-k}\right)^2+k\Delta(x,z,\cos\delta).
\end{align}
One more simplification is a consequence of integration over $\phi$, see Eq. (\ref{reboxintphi}). We obtain the constraint $d(z,\cos\delta)\leq0$ that means $D_0>0$, i.e. the trinomial $R_1(x)$ has no zeros.  Then one can prove
\begin{align}
\label{int_SKx_over_R1sqrtR}
\int\,\rmd x\frac{K\,x+S}{R_1(x)\sqrt{\Delta(x,z,\cos\delta)}}
&=\frac{\mathcal{V}}{2\sqrt{D_0}}\,\ln\abs{\frac{u(x+b)+(C-k)d+\sqrt{D_0\Delta(x,z,\cos\delta)}}
{u(x+b)+(C-k)d-\sqrt{D_0\Delta(x,z,\cos\delta)}}}
\nl
&+\mathcal{U}\arctan\frac{(C-k)(x+b)+u}{\sqrt{(C-k)k\Delta(x,z,\cos\delta)}}
\end{align}
with
\begin{align}
\mathcal{V}&=(C-k)\frac{u(S-Kb)-K(C-k)d}{u^2-(C-k)^2d},\quad
\mathcal{U}=\sqrt{\frac{C-k}{k}}\frac{(S-Kb)(C-k)-Ku}{u^2-(C-k)^2d}\,.
\end{align}

Applying Eq. (\ref{int_SKx_over_R1sqrtR}), we find
\begin{align}
\label{xint_R1sqrtR}
&\int\limits_{-1}^1\rmd x\,\frac{K\,x+S}{R_1(x)\sqrt{\Delta(x,z,\cos\delta)}}\,\sgn(z+x\cos\delta)
=
\frac{\mathcal{V}\mathscr{V}}{\sqrt{D_0}}+\mathcal{U}\mathscr{U},
\end{align}
where
\begin{align}
\mathscr{V}&=\frac12\left[\ln\abs{\frac{u(1+z\cos\delta)+(C-k)d+(z+\cos\delta)\sqrt{D_0}}
{u(1+z\cos\delta)+(C-k)d-(z+\cos\delta)\sqrt{D_0}}}
+\ln\abs{\frac{u(1-z\cos\delta)-(C-k)d+(z-\cos\delta)\sqrt{D_0}}
{u(1-z\cos\delta)-(C-k)d-(z-\cos\delta)\sqrt{D_0}}}\right],
\end{align}
\begin{align}
\mathscr{U}&=
\arctan \frac{u+(C-k)(1+z\cos\delta)}{(z+\cos\delta)\sqrt{(C-k)k}}-\arctan \frac{u-(C-k)(1-z\cos\delta)}{(z-\cos\delta)\sqrt{(C-k)k}}.
\end{align}

In a similar way, we find
\begin{align}
\label{int_MNx_over_R1sqrtR_posd}
&\thfunc{d}\int\,\rmd x\frac{K\,x+S}{R_1(x)\sqrt{-\Delta(x,z,\cos\delta)}}
=
\thfunc{d}\thfunc{D_0}\,\frac{\mathcal{V}}{\sqrt{D_0}}\,
\arctan\frac{u(x+b)+\left(C-k\right)d}
{\sqrt{-D_0\Delta(x)}}
\nl
&-\thfunc{d}\thfunc{D_0}\frac{\mathcal{U}}{2}\,
\,\ln\left|\frac{u+(x+b)\left(C-k\right)-\sqrt{-k(C-k)\Delta(x,z,\cos\delta)}}
{u+(x+b)\left(C-k\right)+\sqrt{-k(C-k)\Delta(x,z,\cos\delta)}}\right|
\nl
&+\thfunc{d}\thfunc{-D_0}\frac{C\,}{4\sqrt{dD}}
\left[
\frac{\tau^-\,f^--\tau^+}{\sqrt{|f^-|}}\,
\ln\left|\frac{(x-x^-)\sqrt{|f^-|}-\sqrt{-\Delta(x,z,\cos\delta)}}{(x-x^-)\sqrt{|f^-|}+\sqrt{-\Delta(x,z,\cos\delta)}}\right|\right.
\nl
&\left.-\frac{\tau^-\,f^+-\tau^+}{\sqrt{|f^+|}}
\ln\left|\frac{(x-x^-)\sqrt{|f^+|}-\sqrt{-\Delta(x,z,\cos\delta)}}{(x-x^-)\sqrt{|f^+|}+\sqrt{-\Delta(x,z,\cos\delta)}}\right|
\right],
\end{align}
where
\begin{align}
f^\pm=-\left[\frac{\sqrt{k(C-k)\,d}\mp\sqrt{|D_0|}}{u-(C-k)\,\sqrt{d}}\right]^2,\quad \sqrt{|f^\pm|}=\frac{\sqrt{-\Delta(X^\pm)}}{X^\pm-x^-},
\end{align}
and $X^\pm$ are the roots of $R_1(x)$. Equation (\ref{R1Deltalink}) implies that $\Delta\left(X^\pm,z,\cos\delta\right)<0$.

We have to consider the integral
$$
\int\,\rmd x\frac{|K\,x+S|}{|R_1(x)|\sqrt{-\Delta(x,z,\cos\delta)}}
$$
together with Eq. (\ref{int_MNx_over_R1sqrtR_posd}). One can show that if $D_0$ is negative, there is no integral of the amplitude due-to additional poles at $x=X^\pm$. It is the reason why we do not need expressions for $\tau^\pm$.

Keeping in mind the property
$$
\thfunc{-\Delta(x,z,\cos\delta)}=\thfunc{1-z^2}\thfunc{x^+-x}\thfunc{x-x^-},
$$
and that the integrand has no additional poles for positive $D_0$, we see that the second term in r.h.s. of Eq. (\ref{int_MNx_over_R1sqrtR_posd}) vanishes. To find values of the first term at $x^\pm$, one should observe that $\abs{u}\geq\sqrt{C(C-k)d}\geq(C-k)\sqrt{d}$  when $D_0>0$, and $\abs{x^\pm+b}=\sqrt{d}$. As a result, we can see that
\begin{equation}
\thfunc{d}\thfunc{D_0}\sgn\left[u(x^\pm+b)+(C-k)d\right]=\pm\thfunc{d}\thfunc{D_0}\sgn\, u.
\end{equation}
Combining all these remarks, one can perform integration over $x$ only when $D_0>0$ with the help of
\begin{align}
\label{xint_R1sqrtR_m}
&\int\limits_{-1}^1\rmd x\,\frac{K\,x+S}{R_1(x)\sqrt{-\Delta(x,z,\cos\delta)}}\,\thfunc{-\Delta(x,z,\cos\delta)}
=\pi\mathcal{V}\frac{\thfunc{1-z^2}}{\sqrt{D_0}}\,\sgn\, u.
\end{align}

Let us take one more step to get expressions that can be directly applied in further evaluations. Namely, we shall deal with integrals like
\begin{align}
\label{intR1sqrtRdecomp}
\int\,\rmd x\frac{Mx+N}{(x-x_0)R_1(x)\sqrt{\pm \Delta(x,z,\cos\delta)}}
=\frac1{R_1(x_0)}\left[Q\int\frac{\rmd x}{(x-x_0)\sqrt{\pm \Delta(x,z,\cos\delta)}}+\int\,\rmd x\frac{S-CQ\,x}{R_1(x)\sqrt{\pm \Delta(x,z,\cos\delta)}}\right]
\end{align}
with $Q=Mx_0+N$, $S=MA-2NB-NCx_0$. The integral with $\sqrt{-\Delta(x,z,\cos\delta)}$ in the denominator is taken using Eqs.~(\ref{xint_xsqrtR_m}) and  (\ref{xint_R1sqrtR_m}). The result is
\begin{align}
\label{xint_MNx_xR1sqrtR_omegacase_m}
\int\limits_{-1}^{1}\rmd x\frac{M\,x+N}{(x-x_0)R_1(x)\sqrt{-\Delta(x,z,\cos\delta)}}\,&\thfunc{-\Delta(x,z,\cos\delta)}
=\frac{\pi\,\thfunc{1-z^2}}{R_1(x_0)}
\nl
&\quad\times\left[\sgn(u)\frac{\mathcal{V}'}{\sqrt{D_0}}
+\sgn(-x_0-z\cos\delta)\frac{Q}{\sqrt{\Delta_0(x_0,z,\cos\delta)}}\right], \quad D_0,\Delta_0>0.
\end{align}

When we deal with $\sqrt{\Delta(x,z,\cos\delta)}$, the second term is evaluated with the help of Eq. (\ref{xint_R1sqrtR}); meanwhile, the first term is just Eq. (\ref{Klevintcorrect}). We derive
\begin{align}
\label{xint_MNx_xR1sqrtR_omegacase}
&\int\limits_{-1}^{1}\rmd x\frac{Mx+N}{(x-x_0)R_1(x)\sqrt{\Delta(x,z,\cos\delta)}}\,\sgn(z+x\cos\delta)
=\frac{1}{R_1(x_0)}\left[\frac{Q\,\Xi(x_0,z,\cos\delta)}{\sqrt{\Delta_0(x_0,z,\cos\delta)}}
+\frac{\mathcal{V}'\mathscr{V}}{\sqrt{D_0}}\right],\quad 1-z^2\leq0,\ 1-x_0^2\leq0,
\end{align}
where
\begin{align}
\mathcal{V}'&=(C-k)\frac{u(S+CQb)+CQ(C-k)d}{u^2-(C-k)^2d}\,.
\end{align}

\subsection{The real and imaginary parts of the integral with four fermion lines}

The expressions obtained in the previous paragraph are sufficient to our purpose. To move further, we need to  know $A$, $B$, $C$, $M$, and $N$.

When we make $x$-integration, $\Psi_{23}(x)$ plays the role of $R_1(x)$. Taking into account the property
\begin{align}
\alpha_{ls}^2+\sin^2\delta_s\sin^2(\phi_l-\phi_s)=\sin^2\psi_{ls},
\end{align}
we find
\begin{align}
C=\sin^2\psi_{23},\quad D=-\sin^2\delta_2\sin^2\delta_3\sin^2\phi_3\Delta_{23}.
\end{align}
Other quantities in Eqs. (\ref{xint_MNx_xR1sqrtR_omegacase_m}) and (\ref{xint_MNx_xR1sqrtR_omegacase}), like $k$, $D_0$, $u$, are not $(l\leftrightarrow s)$ symmetric since they depend on $b$. For integrals with $\Delta_l(x)$, $l=2,3$, we have
\begin{align}
k_l&=\sin^2\delta_s\sin^2(\phi_l-\phi_s),\quad D_{0l}=\alpha_{ls}^2\sin^2\delta_l\Delta_{ls},\quad u_l=(\beta_{ls}-z_l\alpha_{ls}\cos\delta_l)\alpha_{ls},
\end{align}
and
\begin{align}
M_l&=\alpha_{ls}, \quad N_l=\beta_{ls}.
\end{align}
These expressions allow us to find
\begin{align}
\mathcal{V}_l'&=\alpha_{ls}T_1,
\end{align}
where we used that $u=(N-Mb)M$ and $C-k_l=M^2$ which result in $\mathcal{V}'=-M(B+Cx_0)$, and we also introduced the quantities
\begin{equation}
T_l=z_l\sin^2\psi_{st}-z_s(\cos\psi_{ls}-\cos\psi_{lt}\cos\psi_{st})-z_t(\cos\psi_{lt}-\cos\psi_{ls}\cos\psi_{st}),\quad t\neq l,s.
\end{equation}
To simplify the result, we shall also use the following relations:
\begin{equation}
\alpha_{ls}\sin\delta_l=\cos\delta_s-\cos\delta_l\cos\psi_{ls}
\end{equation}
and
\begin{align}
\beta_{ls}-z_l\alpha_{ls}\cos\delta_l&=(z_s-z_l\cos\psi_{ls})\sin\delta_l.
\end{align}

Taking into account all the mentioned properties and applying Eqs. (\ref{xint_delta}), (\ref{xint_MNx_xR1sqrtR_omegacase_m}), and (\ref{xint_MNx_xR1sqrtR_omegacase}), we obtain
\begin{align}
\rp Z_3^\pm&=\frac{1}{4 }\left[ I_{K;12}\find{3}+I_{K;23}\find{3}+I_{K;31}\find{3}\right],\quad \Phi_l(E)\leq0,
\\
\ip Z_3^\pm&=\frac{\pi}{4 }\left[I_{L;12}\find{3}+I_{L;23}\find{3}+I_{L;31}\find{3}\right],\quad \Phi_{ls}(E)>0,
\end{align}
as in the planar case, and $\Phi_{123}(E)\neq0$ when $E\in[m,\Lambda_E]$.
The separated terms $\mathscr{V}_{2,3}$ were joined together using the property
$$
\thfunc{\Delta_{ls}}\left[\sgn(\alpha_{ls})\mathscr{V}_l+\sgn(\alpha_{sl})\mathscr{V}_s\right]=\Xi_{ls}\,\thfunc{\Delta_{ls}}.
$$
At last, one should take into account the relation
\begin{align*}
pk_1k_2k_3k_sk_tT_l&=\pm\left[\vc{b}\find{3}E-\vc{a}\find{3}\right](\vecprod{\vk_s}{\vk_t})
\end{align*}
to reduce the formulas to the form similar to the ``planar'' configuration.


\section{Some words about calculations of integrals with $L+1$ lines}
\label{SecLplus1}

In Sections III-VI, we used Eq. (\ref{Pdelta}) to expand the product under the integral. As one can explicitly check for three line and box integrals, all terms containing two or more $\delta$-functions vanish when the real or imaginary part exists. Generally, it is more convenient to use the only small parameter $\epsilon$ for all poles. In such a way,  a simpler expression can be obtained for any $L$. Namely, transforming the product into a sum with the help of Eq.~(\ref{decomposition}), applying Eq.~(\ref{elimit}), and making inverse transformation to the product, one can find
\begin{align}
\label{Pdelta1}
\prod_{l=1}^L\frac{1}{\lambda_l(E-E_{0l}-i\epsilon)+\vc{p}\vc{k}_l}
&=
\Vp\prod_{l=1}^L\frac{1}{\lambda_l(E-E_{0l})+\vc{p}\vc{k}_l}
\nl
&+i\pi\sum_{l=1}^L\sgn(\lambda_l)\,\delta(\lambda_l(E-E_{0l})+\vc{p}\vc{k}_l)\prod_{s=1,s\neq l}^L\frac{1}{\lambda_s(E-E_{0s})+\vc{p}\vc{k}_s}.
\end{align}
As a result,
\begin{align}
\label{RegZgeneral}
\rp\mathcal{Z}^\pm_L\left(\gX^+_1,\ldots,\gX_L^+;\gT\right)&=\frac{1}{2^L\pi}\,\Vp
\intE p\,f(\pm E-\mu)
\int\limits_{-1}^1\rmd x\int\limits_{0}^{2\pi}\rmd\phi\prod_{l=1}^L\frac{1}{\lambda_l(E-E_{0l})+\vc{p}\vc{k}_l},
\\
\label{ImgZgeneral0}
\ip\mathcal{Z}^\pm_L\left(\gX^+_1,\ldots,\gX_L^+;\gT\right)&=\frac{1}{2^L}\sum_{l=1}^L\sgn(\lambda_l)
\intE p\,f(\pm E-\mu)
\nl
&\times\int\limits_{-1}^1\rmd x\int\limits_{0}^{2\pi}\rmd\phi\,\delta(\lambda_l(E-E_{0l})+\vc{p}\vc{k}_l)\prod_{s=1,s\neq l}^L\frac{1}{\lambda_s(E-E_{0s})+\vc{p}\vc{k}_s},
\end{align}
where we can see only the terms without $\delta$-functions or with a single one.

Equation (\ref{ImgZgeneral0}) can be simplified by performing integration in the imaginary part:
\begin{align}
\label{ImgZgeneral}
\ip\mathcal{Z}^\pm_L\left(\gX^+_1,\ldots,\gX_L^+;\gT\right)&=\frac{1}{2^L}\frac1{{\prod_{s=1}^Lk_s}}\sum_{l=1}^L\sgn(\lambda_l)
\intE \frac{f(\pm E-\mu)}{p^{L-1}}\,\Pi_l(E)\Theta\left(\Phi_l(E)\right),
\end{align}
with
\begin{align}
\label{Pi1def}
\Pi_1(E)&=\int\limits_{0}^{2\pi}\rmd\phi\,\prod_{s=2}^L\frac{1}{z_s-z_1\cos\delta_s+\cos(\phi-\phi_s)\sin\delta_s\sqrt{1-z_1^2}},
\end{align}
and
\begin{align}
\Pi_l(E)
&=\int\limits_{-1}^1\rmd x\frac{\Theta(-\Delta_l(x))}{(z_1+x)\sqrt{-\Delta_l(x)}}\prod_{s=2,s\neq l}^L\frac{\sin\delta_l}{\Psi_{ls}(x)}
\nl
&\times\left[\prod_{s=2,s\neq l}^L[\beta_{ls}+x\alpha_{ls}
-\sin(\phi_l-\phi_s)\sin\delta_s\sqrt{-\Delta_l(x)}]
+\prod_{s=2,s\neq l}^L[\beta_{ls}+x\alpha_{ls}
+\sin(\phi_l-\phi_s)\sin\delta_s\sqrt{-\Delta_l(x)}]\right]
\end{align}
for $l>1$.

The real part can be considered only when $\Phi_l(E)\leq0$, as follows from Sec. IV.B. We can also see that the integrand in the definition of $\Pi_1(E)$ has no singularities when $\Delta_{1s}>0$. For symmetry reasons, one should conclude that the imaginary part exists only if $\Phi_{ls}(E)>0$.

\section{Conclusions}

In the article, we have given the formulas required for numerical evaluation of the four fermion line integral at finite temperature/density that is often needed for theoretical calculations such as within the NJL model. Both the real and imaginary parts of these functions are presented separately. It was proved that the integrals do not exist for some values of the parameters. We give the corresponding constraints for all considered cases.

Derivation of the necessary expressions obliged us to correct the results obtained in \cite{Klevansky} for the integral with three lines and, moreover, to refine the approach used in the cited paper. It was shown that careful consideration of the conditions for the existence plays a crucial role in obtaining valid results.

We have also found, as a by-product, expressions for integrals with any number of lines when momenta are in simple {\it'dot'} or '{\it line}' configurations. For a general configuration, the corresponding expressions are presented as triple and double integrals allowing numerical evaluation.

\vspace{3mm}
  \begin{center} {\bf Acknowledgements:}
We are grateful to A.~Friesen for the idea of this paper and permanent help.
My work was supported by the RFBR grant, no. 18-02-40137, and the NARD project, no. 20.80009.5007.07.

\end{center}

\appendix

\section{Summation of inverse trigonometric functions}
In this Appendix we just cite some beautiful and useful formulas from \cite{Gradshteyn}. The sum and difference of two arccosines are
\begin{align}
\label{arccossum}
\arccos X+\arccos Y&=2\pi\thfunc{-X-Y}+\sgn(X+Y)\arccos\left(XY-\sqrt{1-X^2}\sqrt{1-Y^2}\right),
\\
\arccos X-\arccos Y&=\sgn(Y-X)\arccos\left(XY+\sqrt{1-X^2}\sqrt{1-Y^2}\right).
\end{align}

For arctangents we have
\begin{align}
\label{arctansum}
\arctan X +\arctan Y&=\arctan\frac{X+Y}{1-XY}+\pi\,\sgn(X+Y)\thfunc{XY-1},
\nl
&=\thfunc{|X+Y|}\arctan\frac{XY-1}{X+Y}+\frac\pi2\,\sgn(X+Y),
\end{align}
where the second line is obtained with the help of
\begin{align}
\label{arctan1X}
\arctan X+\arctan \frac1X&=\frac\pi2\,\sgn X.
\end{align}

\section{Some properties of quadratic trinomials arising during the evaluation of integrals}

When we make integration over angles, it is convenient to exclude the variable $E$ using $z_l$. But after that, in the final expressions, we should make the inverse transition as a result of which the following functions arise:
\begin{align}
\Phi_l(E)&\equiv\Phi(E;k_l,\lambda_l,\varepsilon_l,m)=p^2k_l^2(1-z_l^2),
\\
\Phi_{ls}(E)&=p^2k_l^2k_s^2\Delta_{ls},
\\
\Phi_{123}(E)&=p^2k_1^2k_2^2k_3^2\Delta_{123}.
\end{align}
Recognizing that these functions are just trinomials, let us introduce the generalized trinomial
\begin{align}
\Omega_L(E;\gX^+_1,\ldots,\gX^+_L,m)&=C_\Omega(\gX^+_1,\ldots,\gX^+_L,m)\, E^2-2B_\Omega(\gX^+_1,\ldots,\gX^+_L,m) \, E+A_\Omega(\gX^+_1,\ldots,\gX^+_L,m)
\nl
&=\left[\vc{b}\find{L}E-\vc{a}\find{L}\right]^2-\left[w\find{L}\right]^2p^2,
\end{align}
where we demand $\abs{C_\Omega}\neq0$. The coefficients are
\begin{align}
A_\Omega&=\left[\vc{a}\find{L}\right]^2+m^2\left[w\find{L}\right]^2,\quad B_\Omega=\vc{a}\find{L}\vc{b}\find{L},\quad C_\Omega=\left[\vc{b}\find{L}\right]^2-\left[w\find{L}\right]^2.
\end{align}
As we can easily see, the determinant is
\begin{align}
\label{DOmega}
D_\Omega=\left[w\find{L}\right]^2\left\{A_\Omega-m^2\left[\vc{b}\find{L}\right]^2-\frac{\left[\vecprod{\vc{a}\find{L}}{\vc{b}\find{L}}\right]^2}{[w\find{L}]^2}\right\},
\end{align}
where we used that
\begin{align}
\label{vecscalarsum}
\vc{a}^2\vc{b}^2&=\left(\vecprod{\vc{a}}{\vc{b}}\right)^2+\left(\vc{a}\vc{b}\right)^2.
\end{align}

Using the above properties, one can prove that the magnitude of both roots of $\Omega_L(E)$ is not less than $m$. As we have seen, the conditons for the existence of the considered improper integrals have the form $\Omega_L(E)>0$ for any $E\in[ m,\Lambda_E]$. It means that the sign of $\Omega_L(E)$ in the integration interval is fixed which is possible only if both roots lie outside. Then we can prove
\begin{align}
\label{posOmega}
\thfunc{\Omega_L(E)}&=\thfunc{C_\Omega}\thfunc{\Omega_L(E)}
\end{align}
and
\begin{align}
\label{negOmega}
\thfunc{-\Omega_L(E)}&=\thfunc{D_\Omega}\thfunc{-C_\Omega}\thfunc{-\Omega_L(E)}.
\end{align}

The vectors $\vc{a}^{(L)}$ and $\vc{b}^{(L)}$ can be expressed as linear combinations of a set of vectors $\vc{q}_l$ for any $L$:
\begin{align}
\vc{a}\find{L}&=\sum_{l=1}^L\varepsilon_l\vc{q}_l,\quad\vc{b}\find{L}=\sum_{l=1}^L\lambda_l\vc{q}_l,
\end{align}
and $\vc{q}_l=\vc{q}_l\left(\vc{k}_1,\ldots,\vc{k}_L\right)$.

Using the generalized trinomial, one can write
\begin{align}
\label{Phi1def}
\Phi_l(E)&=-\Omega_1\left(E;\gX_l^+,m\right)
\end{align}
with
\begin{align}
\vc{a}\find{1}_l&=\varepsilon_l\frac{\vk}{k},\quad \vc{b}\find{1}_l=\lambda_l\frac{\vk}{k},\quad w\find{1}_l=k_l,
\end{align}
and $\vc{k}$ is an arbitrary vector.

For $L=2$, we have
\begin{align}
\label{Phi2def}
\Phi_{ls}(E)&=\Omega_2\left(E;\gX_l^+,\gX_s^+,m\right),
\end{align}
\begin{align}
\vc{a}\find{2}_{ls}&=\varepsilon_s\vk_l-\varepsilon_l\vk_s,\quad \vc{b}\find{2}_{ls}=\lambda_s\vk_l-\lambda_l\vk_s,\quad w\find{2}_{ls}=\abs{\vecprod{\vk_l}{\vk_s}}.
\end{align}
We also need the coefficients for $L=3$:
\begin{align}
\vc{a}\find{3}&=\varepsilon_1\vecprod{\vk_2}{\vk_3}+\varepsilon_2\vecprod{\vk_3}{\vk_1}+\varepsilon_3\vecprod{\vk_1}{\vk_2},
\nl
\vc{b}\find{3}&=\lambda_1\vecprod{\vk_2}{\vk_3}+\lambda_2\vecprod{\vk_3}{\vk_1}+\lambda_3\vecprod{\vk_1}{\vk_2},
\\
w\find{3}&=\vk_1(\vecprod{\vk_2}{\vk_3}).
\end{align}
One can easily prove
\begin{align}
\vecprod{\vc{a}_{ls}\find{2}}{\vc{b}_{ls}\find{2}}&=\varrho_{ls}(\vk_l\times\vk_s),
\\
\vecprod{\vc{a}\find{3}}{\vc{b}\find{3}}&=
w\find{3}\left(\varrho_{12}\vk_3-\varrho_{13}\vk_2+\varrho_{23}\vk_1\right),
\end{align}
where
\begin{align}
\varrho_{ls}&=\lambda_s\,\varepsilon_l-\lambda_l\varepsilon_s=\lambda_l\lambda_s(E_{0l}-E_{0s}).
\end{align}
Then, using
\begin{align}
\vc{a}\find{3}&=\varepsilon_t(\vecprod{\vk_l}{\vk_s})-\vecprod{\vc{a}\find{2}_{ls}}{\vk_t},
\nl
\vc{b}\find{3}&=\lambda_t(\vecprod{\vk_l}{\vk_s})-\vecprod{\vc{b}\find{2}_{ls}}{\vk_t},
\end{align}
and Eq. (\ref{vecscalarsum}), we find
\begin{align}
\label{Phi123Phils}
\Phi_{123}(E)\left(\vecprod{\vk_l}{\vk_s}\right)^2&=\left\{\left[\vc{b}\find{3}E-\vc{a}\find{3}\right]\left(\vecprod{\vk_l}{\vk_s}\right)\right\}^2
+\left[w\find{3}\right]^2\Phi_{ls}(E)
\end{align}
from which one can see that $\Phi_{123}(E)\geq0$ if $\Phi_{ls}(E)\geq0$.

Similar expressions are also valid for $\Phi_{ls}(E)$. The relation
\begin{align}
\label{PhilsPhil}
\Phi_{ls}(E)k_s^2&=\left\{\left[\vc{b}\find{2}E-\vc{a}\find{2}\right]\vk_s\right\}^2-\left[w\find{2}\right]^2\Phi_l(E)
\end{align}
results in non-negativity of $\Phi_{ls}(E)\geq0$ if $\Phi_l(E)\leq0$.

The generalized trinomial $\Omega_L$ allows one to deal in the same way with the constraints on $k_l$ and integrands containing $\thfunc{\pm\Omega_L}$ or $\Omega_L^{-1}$ like $I_{H,K,L}^\pm$.

\end{document}